\documentclass [10pt]{article} 

\begin{document} 

\title{Time in Quantum Geometrodynamics} 
\author{Arkady Kheyfets\thanks{E-mail: kheyfets@math.ncsu.edu}\\
Department of Mathematics\\ 
North Carolina State University\\ 
Raleigh, NC 27695-8205
\and 
Warner A.~Miller\thanks{E-mail: wam@lanl.gov}\\
Theoretical Astrophysics Group (T-6, MS B288)\\ 
Theoretical Division\\ 
Los Alamos National Laboratory\\
Los Alamos, NM 87545}
\date{\today}

\maketitle 

\begin{abstract} 
We revisit the issue of time in quantum geometrodynamics and suggest 
a quantization procedure on the space of true dynamic variables. This 
procedure separates the issue of quantization from enforcing the 
constraints caused by the general covariance symmetries. The resulting 
theory, unlike the standard approach, takes into account the states that 
are off shell with respect to the constraints, and thus avoids the problems 
of time. In this approach, quantum geometrodynamics, general covariance, and 
the interpretation of time emerge together as parts of the solution of the 
total problem of geometrodynamic evolution.
\end{abstract}



\section{Introduction.} 
\label{I}

A proper introduction of time is one of the central issues of any viable 
programme of gravity quantization. Its resolution is important conceptually 
as it determines in a profound way the meaning of the quantization procedure 
or the meaning of the basic structure that contains quantum gravity as 
a particular case or an approximation. It is also likely that a proper 
understanding of this issue will provide the clues to answering the questions 
motivating gravity quantization in the first place, such as avoidance of 
singularities, the issues of the final and the initial state of the Universe, 
etc. It might contribute to a better understanding of the aspects of modern 
quantum field theories that depend crucially on the causal structure of 
spacetime. One can be referred to the recent paper \cite{Isham99} by 
J.~Butterfield and C.~J.~Isham for an exhaustive exposition of motivational 
and conceptual problems of quantum gravity, as well as for an overview of 
modern programmes of gravity quantization.  

Quite understandably, the issue of time is formulated differently in 
different approaches to gravity quantization. In what follows we restrict 
ourselves to what can be thought of as a slight extension of the canonical 
gravity quantization programme. 

The standard canonical quantum gravity approach \cite{Kuc93} is based on the 
classical dynamic picture of the evolving 3--geometry of a slicing of 
a spacetime manifold. The slicing is essentially a reference foliation of the 
spacetime manifold (endowed with a 4--geometry) with respect to which 
the canonical variables are assigned. It is usually parametrized by the 
lapse function $N$ and the shift functions $N^i$. The canonical variables 
are the 3--metric $g_{ik}$ on a spatial slice $\Sigma$ of the foliation 
induced by the spacetime 4--metric, and their canonical conjugate matrix 
$\pi^{ik}$. The latter is related to the extrinsic curvature of $\Sigma$ when 
it is considered as embedded in the spacetime.  

The customary variational procedure applied to the Hilbert action expressed 
in terms of the canonical variables produces Hamilton equations  
describing the time evolution of the canonical variables, with the 
Hamiltonian given as $N {\cal H} + N^i {\cal H}_i$ where ${\cal H}$ and 
${\cal H}_i$ are functions of the canonical variables and their spatial 
derivatives. The procedure is not extended to the derivation of the 
Hamilton--Jacobi equation in the usual manner as such an equation is 
rendered to be meaningless with the chosen set of canonical variables 
(cf.~\cite{Mis57}) when the general covariance of the theory is taken into 
account.    

As a way out, the general covariance is introduced in the variational 
principle from the very onset as the requirement of the action to be 
invariant with respect to variations of the lapse and shift which leads 
to the constraint equations (to simplify notations, we omit indices on 
components of $g$ and $\pi$ in all equations below) 
\begin{equation} 
{\cal H}(g, \pi ; x) = 0,  
\end{equation}   
and 
\begin{equation} 
{\cal H}_i(g, \pi ; x) = 0,   
\end{equation}   
imposed on the canonical variables on each slice. An important feature 
of general relativity is that its dynamics is fully constrained. It can be 
shown that if the geometry of spacetime is such that the constraints are 
satisfied on all the slices of all spatial foliations of spacetime, then 
the canonical variables necessarily satisfy the Hamilton evolution equations. 
This feature is often referred to as a key property of general relativity 
\cite{Isham99} and is interpreted as an argument that the entire theory is 
coded in the constraints, with the conclusion that the Hamilton equations are 
redundant and can be ignored in dynamic considerations. Substitution of  
$\delta S/\delta g$ in the place of $p$ in constraint equations leads to 
the new set of equations 
\begin{equation} 
{\cal H}\left( g, {\delta S\over\delta g}; x\right) = 0,  
\end{equation}   
and 
\begin{equation} 
{\cal H}_i\left( g, {\delta S\over\delta g}; x\right) = 0,   
\end{equation}  
the first of which is considered to be the Hamilton--Jacobi equation. This 
assertion is supported by arguments appealing to the variational principle on 
superspace of 3--geometries (for detailed arguments and the interpretation of 
other equations cf. \cite{MTW70}). 

Dirac's procedure of canonical gravity quantization is based directly on this 
Hamilton--Jacobi equation and produces the quantum theory that consists of 
commutation relations imposed on all canonical variables and the 
Wheeler -- DeWitt equation. 

The ADM square root quantization procedure is also based entirely on 
constraints, but in this procedure the set of canonical variables is split in 
two subsets, embedding variables (four of them altogether; one slicing 
parameter $\Omega$ and three coordinatization parameters $\alpha$) and 
true dynamic variables $\beta$ (two of them) \cite{Kuc92}, \cite{KheMil94a}, 
\cite{KheMil96}. The set of constraints is 
solved with respect to the momenta conjugate to the embedding variables. 
After substituting $\delta S/\delta\Omega$, $\delta S/\delta\alpha$ for 
$p_\Omega$, $p_\alpha$, (where $S$ is the principal Hamilton function) 
 one of the resulting equations 
(the equation for the momentum conjugate to the slicing parameter) is 
identified with the Hamilton--Jacobi equation, and its right hand side 
yields an expression for a new (square root) Hamiltonian. The quantization 
is based on this equation and produces the quantum theory that consists of 
the Schr\" odinger equation and commutation relations imposed on true dynamic 
variables and their conjugate momenta. 

In both approaches, the description of time evolution of quantized 
gravitational fields or systems including such fields becomes extremely 
troublesome. Any attempt to introduce time that can be used in a way similar to 
that of time in quantum mechanics or in quantum field theory on a flat 
background invariably leads to the notorious problems of time \cite{Kuc92}, 
some of which are of a conceptual nature while others are technical. Attempts 
to introduce time in such systems in a universal way from outside, as a 
reading of a specially designed clock have been unsuccessful and there are all 
reasons to believe that it is impossible \cite{Isham99}, whether the clock is 
believed to be gravitational (the readings depend only on the variables 
describing gravitational field) in its nature or it is a matter clock, for as 
long as it interacts with gravity.  

The difficulties of the conceptual nature (the problem of 
functional evolution, and the multiple choice problem, in Kucha\v r's 
terminology) emerge due to the dual nature of time parametrization in 
general relativity. If spacetime is considered as a manifold it can be 
coordinatized and sliced in any arbitrary manner. However, this is not 
sufficient for the description of geometrodynamic evolution. Both slicing and 
coordinatization need to be tied to the metric on the manifold. The 
standard way of doing it in classical geometrodynamics involves 
lapse and shift. These parameters express slicing condition in 
terms of readings of the clocks of resting test particles, and 
coordinatization conditions as the metric shift of the coordinate grid 
on the slice with time. In such a description, the reduction of the dynamic 
picture to the constraints is based entirely on existence of the 
unique spacetime metric (although the metric might be not known until 
the geometrodynamic problem is resolved).  
While this is not a problem in classical general relativity, there is, 
in general, no possibility to assign such a unique metric on spacetime in 
canonical quantum gravity. 

With this in mind, keeping the constraints equations as a foundation of 
geometrodynamics becomes not very meaningful, to say the least. Even in the 
classical theory, such a procedure results in the treatment of the 
Hamilton--Jacobi equation, that looks quite artificial and totally different 
from the standard mechanical considerations. From the point of view of 
Hamilton dynamics, shift and lapse invariance is just a symmetry of the 
system and should not be used in deriving dynamic equations. 

In fact, the derivation of Hamilton equations does not 
depend on assumptions of lapse and shift invariance of action (general 
covariance). It is only in the derivation of the Hamilton--Jacobi equation 
this invariance becomes involved in a fundamental way and essentially 
replaces standard dynamic considerations. 

As a result, the Hamiltonian of Hamilton equations does not coincide with the 
Hamiltonian of the Hamilton--Jacobi equation, only the first being related to 
the Lagrangian in a standard way. Also, the Hamilton--Jacobi equation does 
not contain the reference to the time evolution at all. 

Quantization of the dynamic picture based on the constraints is essentially 
equivalent to restricting the states of the resulting quantum systems to 
a ``shell'' determined by the constraints that are classical in their origin. 
An attempt to undertake a similar action in quantum mechanics or quantum 
field theory would be quite disastrous under all but very carefully 
selected conditions. 

One way to resolve this dilemma would be to weaken the requirement of  
covariance, essentially discarding it in dynamic considerations and 
recovering it by imposing symmetries on solutions only to the extent and 
in the sense that is allowed by dynamics. The general covariance in its 
traditional meaning should be recovered in the classical limit. In 
a sense, this requirement should determine, at least partially, what 
constitutes the classical limit in quantum geometrodynamics. 
  
In order to achieve this goal a formulation of geometrodynamic 
Hamilton--Jacobi theory independent of the symmetry assumptions is needed, 
in the same spirit as in a standard setting of mechanics.  
Is there a possibility of writing the Hamilton--Jacobi equation in a way 
that is closer to that encountered in mechanics? As we have mentioned, 
the answer is no, if one considers as the object of geometrodynamics the 
whole 3--metric (or 3--geometry) of the slice \cite{Mis57}. 

The situation changes dramatically if York's analysis of gravitational 
degrees of freedom \cite{Yor72} is taken into account and actively utilized. 
It becomes possible to reformulate classical geometrodynamics in a standard 
way from the very beginning to the very end and to treat general covariance 
as a symmetry of gravitational systems. Although the Hamilton--Jacobi 
equation looks different and makes sense in its traditional form, the 
resulting description is equivalent to the commonly accepted in classical 
geometrodynamics. However, quantization based on this new Hamilton--Jacobi 
equation provides an appropriate interpretation of the conceptual problems 
of time making them quite natural statements concerning the properties of 
gravity quantization. It also seems to avoid the technical problems of time, 
such as the Hilbert space problem, and the spectral analysis problem, as it 
produces the Schr\" odinger equation for the state evolution, and the 
Hamiltonian does not include the square root operation. The procedure has 
been described previously elsewhere \cite{KheMil94a}, \cite{KheMil96}, but 
we do not believe that it is widely known and provide a brief overview of it 
below. 

In this setting, time can be introduced as a slicing parametrization on 
the spacetime manifold and tied to the metric structure without any 
contradictions. The metric interpretation of time is coupled with 
geometrodynamic evolution. The true meaning of time becomes completely 
determined only after the geometrodynamic evolution problem has been solved. 
In a sense, quantum geometrodynamic configuration and time emerge 
together and the meaning of the clock readings is influenced by quantum 
gravitational system. 

We illustrate the emerging meaning of time by considering different examples 
and introducing times as counted by matter clocks and by gravitational 
clocks. These clocks are akin to the clocks measuring timelike intervals 
along the world lines of test particles (matter clock) and the York's 
extrinsic time parameter. In the end, we discuss the general features of 
time parametrization and its metric interpretation in quantum 
geometrodynamics. 


\section{Geometrodynamic Quantization in General \newline Setting.}
\label{II}

According to York's analysis of gravitational degrees of freedom, 
the set of six parameters describing the slice 
3--metric should be split in two subsets, $\{\beta_1, \beta_2\}$ (two 
functions) and $\{\alpha_1, \alpha_2, \alpha_3, \Omega\}$. The first of these 
is treated as the set of true gravitational degrees of freedom (the 
initial values for them can be given freely), while the second is considered 
to be the set of embedding variables. The $\alpha$ parameters are often 
referred to as coordinatization parameters, while $\Omega$ is called, 
depending on the context, the slicing parameter, the scale factor, or the 
many--fingered time parameter. Information relevant to the gravity field is 
carried by $\beta$ parameters, while $\alpha$ and $\Omega$ essentially 
describe time. In the original York's analysis the $\beta$ variables describe 
the conformal part of a slice 3--geometry, $\Omega$ represents the scale 
factor and the $\alpha$ variables are determined by the choice of 
coordinatization of 3--slices. The true dynamic 
variables form what we call a dynamic superspace while the 
embedding variables are treated as functional parameters. 

The idea is to develop geometrodynamics from the very beginning on the 
dynamic superspace instead of the superspace of 3--metrics or 3--geometries. 
The variational principle on the dynamic superspace or its phase space 
(formed by true dynamic variables $\{\beta_1, \beta_2\}$ and their conjugate 
momenta $\{\pi_{\beta_1}, \pi_{\beta_2}\}$ yields the dynamic equations 
describing evolution of true dynamic variables. All of these equations 
depend on lapse and shift and contain embedding variables as functional 
parameters. These are treated as an external field and is determined by 
additional equations that do not follow from the variational principle on 
dynamic superspace. The quantization procedure is performed on the dynamic 
superspace (only ${\beta}$-s are quantized, i.~e.~generate commutation 
relations, while embedding variables form a classical field). The 
Schr\" odinger equation is obtained by a quantization procedure from the 
Hamilton--Jacobi equation on the dynamic superspace and describes the time 
evolution of the state functional on true dynamic superspace coupled with the 
external classical field determined by embedding variables. In this paper 
we introduce such a coupling via a procedure similar to that of 
Hartree--Fock. 

In a more detailed and precise description that follows, we omit indices on 
variables $\beta$ and $\alpha$ for the sake of notational simplicity. They can 
be recovered easily whenever it becomes necessary. 

We start from the standard Lagrangian ${\cal L}$ (written in terms of
the 3--metric, shift and lapse) and the associated action (with
appropriate boundary terms, as needed, to remove the terms containing
second time derivatives) and we introduce the momenta conjugate to the
true dynamic variables
\begin{equation} 
\label{301} 
\pi_\beta = {\partial {\cal L}\over \partial\dot\beta}. 
\end{equation} 
We then use these $\pi_\beta$'s to form the geometrodynamic Hamiltonian 
${\cal H}_{dyn}$, 
\begin{equation} 
\label{302} 
{\cal H}_{dyn} = \pi_\beta \dot\beta - {\cal L}. 
\end{equation} 
The arguments of the Hamiltonian ${\cal H}_{dyn}$ are described by 
the expression
\begin{equation} 
\label{303} 
{\cal H}_{dyn} = {\cal H}_{dyn}(\beta , \pi_\beta  ; \Omega ,  
\alpha ). 
\end{equation}
The variables following the semicolon are treated as describing an 
external field,
while the ones preceding the semicolon are the coordinates and
momenta of the gravitational true degrees of freedom, i.e. of the
true geometrodynamics. The variation of $\beta$ and $\pi_\beta$
leads to the equations of geometrodynamics, i.e. to two pairs of
Hamilton equations,
\begin{equation} 
\label{304} 
\dot\beta = {\partial {\cal H}_{dyn}\over\partial\pi_\beta}, 
\end{equation}  
\begin{equation} 
\label{305} 
\dot\pi_\beta = - {\partial {\cal H}_{dyn}\over\partial\beta}, 
\end{equation} 
and, subsequently, to the Hamilton--Jacobi equation 
\begin{equation} 
\label{306}  
{\delta S\over\delta t} = - {\cal H}_{DYN}\left(\beta , 
{\delta S\over\delta\beta}; \Omega ,  
\alpha \right). 
\end{equation} 
Here $S$ is a functional of $\beta$ and, in addition, a function of $t$, 
\begin{equation} 
\label{307} 
S = S\left[\beta ; t \right) . 
\end{equation} 
and ${\delta \over \delta t}$ is defined by  
\begin{equation} 
\label{307a} 
{\partial S \over \partial t} = \int {\delta S\over\delta t} d^3x. 
\end{equation} 

Neither the Hamilton equations (\ref{304}), (\ref{305}) nor the
Hamilton--Jacobi equation (\ref{306}) are capable of providing any
predictions as their solutions depend on the functional parameters
$\Omega$ and $\alpha$ which are not yet known.  One can complete the
system of equations by adding to the Hamilton equations, or to the
Hamilton--Jacobi equation, the standard constraint equations of
general relativity.  They should be satisfied when the solution 
for $\beta$, $\pi_\beta$ of
equations of true geometrodynamics (with appropriate initial 
data) is substituted in them (we use symbols $[\beta ]_s$, $[\pi_\beta ]_s$ 
for such a solution) 
\begin{equation} 
\label{308} 
 \matrix{{\cal H}^i\left( [\beta ]_s, [\pi_\beta ]_s , 
\Omega , \alpha\right) = 0 \cr 
  \cr 
{\cal H}\left( [\beta ]_s, [\pi_\beta ]_s ,  
\Omega , \alpha\right) = 0 \cr}
\end{equation} 
These constraint equations cannot be derived from variational
principles on dynamic superspace. Rather, they should be treated as
additional symmetries, or the equations for an external field. They do
follow from the shift and lapse invariance of the action but their
derivation in this new setting depends on the structure of the whole
action integral (cf.~section \ref{IV}).  As a result, they cannot
replace the full set of equations for geometrodynamic
evolution. However, the resulting complete system of equations
(dynamic equations on conformal superspace and constraint equations)
is equivalent to this of the standard geometrodynamics on the
superspace of 3--geometries \cite{KheMil96}.

For the purpose of quantization, we make a
transition to the corresponding Schr\"odinger equation based entirely 
on dynamics and ignoring the system symmetries 
\begin{equation} 
\label{309} 
i \hbar {\delta\Psi\over\delta t} = \widehat{\cal H}_{dyn}\left(\beta , 
\widehat\pi_\beta ; \Omega , \alpha\right) \Psi 
\end{equation} 
where $\widehat\pi_\beta = {\hbar\over i} {\delta\over\delta\beta}$. 
The Schr\"odinger equation (\ref{309}) implies that commutation relations 
are imposed only on true dynamic variables and treats embedding variables 
as external classical fields. The state 
functional $\Psi$ in this equation is a functional of $\beta$ 
and a function of $t$.  
\begin{equation} 
\label{310} 
\Psi = \Psi\left[\beta , t\right) 
\end{equation} 
This Schr\"odinger equation (with specific initial data) can be solved
(cf., for instance the example of the Bianchi~1A cosmological model
below). The resulting solution $\Psi_s$ of this Schr\"odinger equation
is not capable of providing any definite predictions as it depends on
four functional parameters $\Omega$, $\alpha$ which remain at this
stage undetermined. All expectations, such as the expectation values
of $\beta$
\begin{equation} 
\label{311} 
<\beta >_s = \langle\Psi_s\vert\beta\vert\Psi_s\rangle = 
\int\Psi^*_s \beta \Psi_s \, {\cal D}\beta 
\end{equation}
or of $\widehat\pi_\beta$ 
\begin{equation} 
\label{312} 
<\pi_\beta >_s = \langle\Psi_s\vert\widehat\pi_\beta\vert\Psi_s\rangle = 
\int\Psi^*_s \widehat\pi_\beta \Psi_s \, {\cal D}\beta 
\end{equation} 
also depend on these functional parameters.  To specify these
functions we resort to the constraint equations. The treatment of the
constraints has nothing to do with quantization of geometrodynamics.
It is merely introducing the coupling between already quantized
geometrodynamics and the classical field determined by embedding
variables.  In other words, they take care of the symmetries, which are
classical in their nature, to the extent they are capable of doing
that.

As in case of classical geometrodynamics, we impose the constraints on
the solution of the dynamic equations (Schr\"odinger equation) with
appropriate initial data and in this way, determine the unique values
of $\Omega$ and $\alpha$. It is possible that there are several ways
to couple the constraints to the quantization of the true dynamic
variables, $\beta$.  As per our previous proposal we impose the four
constraints only on the expectation values of the conformal dynamics
\begin{equation} 
\label{313} 
\matrix{{\cal H}^i\left( <\beta >_s, <\pi_\beta >_s , 
\Omega , \alpha\right) = 0 \cr 
  \cr 
{\cal H}\left( <\beta >_s, <\pi_\beta >_s ,  
\Omega , \alpha\right) = 0 \cr}
\end{equation}  

The way evolution occurs can be described as follows. Initial data at 
$t = t_0$ 
consist of the initial state functional $\Psi = \Psi_0$ and the initial 
values (functions) of embedding variables. In addition, lapse and shift are 
supposed to be given. Equations (\ref{311}), (\ref{312}) yield the 
expectation values (functions) of true dynamic variables and their conjugate 
momenta. The result are substituted in the constraints (\ref{313}). After 
this, the constraints are solved with respect to the time derivatives of 
embedding variables. A step forward in time (say, with the increment 
$\Delta t$) is performed by integration of obtained expressions to evolve 
embedding variables and and by integration of the Schr\" odinger equation 
(\ref{309}) to evolve the state functional. This concludes one step forward 
in time. The next step is performed by repeating the same operations in the 
same order. 

One can be referred to \cite{KheMil94a}, \cite{KheMil96} for two particular 
examples 
illustrating such geometrodynamic evolution in cases of Bianchi 1A cosmology 
and Taub cosmology. The first one can and has been solved analytically, while 
the latter one has been solved numerically. 

We provide an abbreviated description of only the first example (Bianchi 1A) 
as its analytical solution is more useful in presenting the issue of time 
in quantum geometrodynamics.

\section{Geometrodynamic Quantization: Bianchi 1A Model.}
\label{III}

The Bianchi 1A cosmological model is commonly referred to as the
axisymmetric Kasner model \cite{RS75}. Its metric is determined by two
parameters, the scale factor $\Omega$ and the anisotropy parameter
$\beta$ 
\begin{equation}
\label{401}
ds^2 = - dt^2 + {\rm e}^{-2\Omega} \left( {\rm e}^{2\beta} dx^2 + 
{\rm e}^{2\beta} dy^2 + {\rm e}^{-4\beta} dz^2 \right).
\end{equation}
The choice of this expression for the metric implies that we have chosen 
$N^i = 0$ and $N =1$ values of shift and lapse for this example.
As this cosmology is homogeneous the two functions $\Omega$ and $\beta$ are 
the functions of the time parameter $t$ only. The scalar 4--curvature can 
be expressed in terms of these two functions to yield the Hilbert action and, 
after subtracting  the boundary term, the cosmological action,
\begin{equation}
\label{402}
I_C = I_H + {3V\over 8\pi}\dot\Omega {\rm e}^{-3\Omega}\vert_{t_0}^{t_f} = 
{3V\over 8\pi}\int\limits_{t_0}^{t_f} \left( \dot\beta^2 - \dot\Omega^2 
\right) {\rm e}^{-3\Omega} dt,
\end{equation}
where $V = \int\int\int dx dy dz$ is the spatial volume element. As it is 
usually done in case of homogeneous cosmologies, we integrate 
appropriate quantities 
over the spatial volumes and work with integrated Lagrangian $L$, Hamiltonian 
$H$ and momenta $p_\beta$ rather than with their densities ${\cal L}, 
{\cal H}, \pi_\beta$.  

We treat the scale factor $\Omega (t)$ as the embedding variable and 
the anisotropy $\beta (t)$ as the dynamic degree of freedom. The momentum 
conjugate to $\beta$ is 
\begin{equation}
\label{403}
p_\beta = {\partial L \over \partial\dot\beta} = {3V\over 4\pi} 
{\rm e}^{-3\Omega} \dot\beta.
\end{equation} 
The Hamiltonian of the system in our approach can be expressed in terms of 
the momentum conjugate to $\beta$ and the Lagrangian.
\begin{equation}
\label{404}
H_{dyn}  =  p_\beta \dot\beta - L \nonumber = 
{2\pi \over 3V} {\rm e}^{3\Omega} p_\beta^2 + {3V\over 8\pi} 
\dot\Omega^2 {\rm e}^{-3\Omega}. 
\end{equation}
In the classical theory this Hamiltonian can be used to produce either one
pair of Hamilton equations or the equivalent Hamilton--Jacobi
equation.  In either case, the dynamics picture derived in this way is
incomplete. To complete it, we impose the super-Hamiltonian
constraint.
\begin{equation}
\label{405}
p_\beta^2 = \left(  {3V \over 4\pi}\right)^2 {\rm e}^{-6\Omega} \dot\Omega^2.
\end{equation}

Using the Hamilton--Jacobi equation,
\begin{equation}
\label{406}
{\partial S\over \partial t} = 
- H_{dyn}\left( {\partial S\over \partial\beta}, 
\Omega (t), \dot\Omega (t) \right),
\end{equation}
together with the expression (\ref{404}) for the Hamiltonian $H_{dyn}$, 
we obtain the Schr\"odinger equation for the axisymmetric Kasner model.
\begin{equation}
\label{407}
i\hbar {\partial\Psi\over\partial t} = -{2\pi\hbar^2 \over 3V} 
{\rm e}^{3\Omega} {\partial^2 \Psi \over \partial\beta^2} + 
{3V\over 8\pi} \dot\Omega^2 {\rm e}^{-3\Omega} \Psi .
\end{equation}
The constant $\hbar$ in this equation should be understood as the square of 
Planck's length scale, rather than the standard Planck constant. The quantum 
picture based on the Schr\"odinger equation (\ref{407}) is incomplete as the 
scale factor $\Omega$ is so far an unknown function of time.  
To complete the dynamics picture we follow our prescription and impose, in 
addition to equation (\ref{407}), the super-Hamiltonian constraint.
\begin{equation}
\label{408}
<p_\beta>_s^2 = \left( {4\pi \over 3V}\right)^2 {\rm e}^{-6\Omega} 
\dot\Omega^2.
\end{equation}
Here $<p_\beta>_s$ is the expectation value of the momentum 
$\widehat p_\beta = 
{\hbar\over i} {\partial \over \partial\beta}$ 
\begin{equation}
\label{409}
<p_\beta>_s = \langle\Psi_s\vert\widehat p_\beta\vert\Psi_s\rangle = 
\int\limits_{-\infty}^{\infty} \Psi_s^*(\beta , t) \widehat p_\beta \Psi_s 
(\beta , t) d\beta 
\end{equation}
where $\Psi_s$ is the solution of the Schr\"odinger equation with 
specified initial data.
The system of equations (\ref{407}), (\ref{408}) provide us with a complete
quantum dynamic picture of the axisymmetric Kasner model evolution and, when
augmented by appropriate initial and boundary conditions, can be
solved analytically.

For instance, we can specify the initial data for the Schr\" odinger 
equation in the form 
\begin{equation}
\label{417}
\Psi (\beta , t)\vert_{t_0} = \Psi_s(\beta , t_0) = 
\int\limits_{-\infty}^{\infty} A_k {\rm e}^{{i\over\hbar}k\beta} dk 
\end{equation}
with 
\begin{equation}
\label{419}
A_k = C {\rm e}^{-a(k - k_0)^2}.
\end{equation}
Such a choice introduces the Gaussian wave 
packet centered initially at the value $k_0$ of $k$ (we will describe the 
meaning of $k_0$ later), and of the initial width determined by the 
constant $a$, with $C$ being merely the normalization constant, picked to 
satisfy the condition 
$\langle\Psi (\beta , t_0)\vert\Psi (\beta , t_0)\rangle  = 1$ . 

The solution of the Schr\" odinger equation with such initial data can be 
written as 
\begin{equation} 
\label{419a} 
\Psi_s(\beta , t) = 
C \sqrt{\pi} \left( a^2 + {f^2\over\hbar^2}\right)^{-{1\over 4}}\, 
{\rm exp}\left\{ -{a\over 4\left( a^2 + {f^2\over\hbar^2}\right) }\, 
{(\beta - 2 k_0 f)^2\over\hbar^2}\right\}\ {\rm e}^{iF}, 
\end{equation} 
where 
\begin{equation}
\label{423}
f = f(t) = {2\pi\over 3V}\int\limits_{t_0}^t {\rm e}^{3\Omega} dt, 
\end{equation} 
and $F = F(\beta , t)$ is rather involved real valued expression 
\cite{KheMil94a} 
that depends on $\beta$, $\Omega$, and $\dot\Omega$. However, it has 
a structure that makes computations of relevant expectation values easy. 
Such a computation of the expectation $\langle p_\beta\rangle_s$ of the 
momentum $\hat p_\beta = {\hbar\over i} {\partial\over\partial\beta}$ yields 
\begin{equation}
\label{428}
<p_\beta>_s = \langle\Psi_s\vert\widehat p_\beta \vert\Psi_s\rangle = k_0. 
\end{equation}
In other words the expectation value of the momentum $<p_\beta >_s$ does not 
change with time. It is determined by the $k$--center of the packet at 
$t=t_0$.
 

It is clear that this solution of the Schr\"odinger equation
describing the wave packet time evolution cannot provide any 
definite predictions as it contains as yet undetermined scale factor $\Omega$.
To find $\Omega(t)$ we need to substitute this expectation value
into the constraint (\ref{408}) and to solve the resulting equation
with respect to $\Omega$. Substitution of $\langle p_\beta\rangle_s = k_0$ 
in (\ref{408}) yields 
\begin{equation}
\label{429}
k_0^2 = \left( {3V \over 4\pi}\right)^2 {\rm e}^{-6\Omega} \dot\Omega^2.
\end{equation}
Once the solution of this equation is substituted in (\ref{419a})
the geometrodynamic problem (\ref{407}), (\ref{408}) for the wave packet
(\ref{419}) is solved completely. 

The solution can be used to compute the expectation value for $\beta$:
\begin{equation}
\label{430}
<\beta >_s = \langle\Psi_s\vert\beta\vert\Psi_s\rangle = 2k_0f(t),   
\end{equation} 
and the variance in $\beta$ 
\begin{equation}
\label{431}
<\left(\beta\  - <\beta >\right)^2>_s = 
\langle\Psi_s\vert \left(\beta\  - <\beta >\right)^2 \vert\Psi_s\rangle = 
{\hbar^2 a^2 + f^2 \over a}
\end{equation} 
Thus ``the center'' of the wave packet evolves as the classical Kasner
universe determined by the momentum value equal to $k_0$ would evolve,
while the spread of the packet increases
with time.  The result is similar to that of the quantum mechanics of
a free particle; after all the Bianchi~I cosmology is the
free--particle analogue of quantum cosmology.

\section{Time in Quantum Geometrodynamics.}
\label{IV} 

In sections 
\ref{II}, \ref{III} we have demonstrated that a proper understanding of 
classical geometrodynamics as the dynamics on the superspace of true 
dynamic variables amended by the constraints attributed to the universal 
symmetries of gravitational systems (lapse and shift invariance) opens 
a way to circumvent the problems of time (for a full discussion, cf. 
\cite{KheMil94a}, \cite{KheMil96}). 

The important point is that these 
problems disappear as soon as the proper object of quantization (true 
geometrodynamic variables) is chosen. The constraints themselves are 
of no primary significance in this process. Their presence in the theory 
reflects the fact that for gravity the true dynamics cannot be decoupled 
from the evolution of ``embedding'' variables and that within the classical 
theory the true dynamic variables picture is limited to a ``shell'' 
determined by the constraints. There is absolutely no reason to expect 
that the last feature will survive after quantization, except, perhaps, for 
some particular carefully chosen systems. We do not require it and thus 
avoid the problems of time. 

The particular choice of lapse ($N = 1$) and shift ($N^i = 0$) in the 
previous section is quite sufficient to make this point. However, such 
a choice becomes an obstacle for understanding the subtle differences 
between our treatment of constraints and the standard one. Also it fixes 
a particular choice of time and its interpretation, thus precluding the study 
of alternatives in the important issue of time in quantum geometrodynamics. 
Essentially, we have become tied to the classical matter clocks at rest as 
determined by spacelike slices of the Kasner universe. We cannot even switch 
to another classical clock, such as the one that produces the trace of the 
extrinsic curvature as its readings, while such a clock might be relevant 
in resolving the issue of the final singularity for some cosmological 
models. Such a transition demands 
releasing of at least lapse $N$ and making it a function of time $N = N(t)$ 
($N = N(K)$, if we want to assign $t = K = {\rm Tr}(K)$). 

This amounts to the choice of metric in the form     
\begin{equation}
\label{501}
ds^2 = -\left[ N(t) dt\right]^2 + {\rm e}^{-2\Omega (t)} 
\left[ {\rm e}^{2\beta (t)} dx^2 + 
{\rm e}^{2\beta (t)} dy^2 + {\rm e}^{-4\beta (t)} dz^2 \right].
\end{equation} 
instead of (\ref{401}) (we still retain zero shift $N^i = 0$), which results 
in the expression for the Lagrangian 
\begin{equation} 
\label{502} 
L = \frac{3v}{8\pi}\, \frac{1}{N}\, \left( \dot{\beta}^2 - 
\dot{\Omega}^2\right)\, {\rm e}^{-3\Omega} 
\end{equation}  
The momentum conjugate to $\beta$ becomes  
\begin{equation}
\label{503}
p_\beta = {\partial L \over \partial\dot\beta} = {3V\over 4\pi}\, {1\over N}\, 
{\rm e}^{-3\Omega} \dot\beta.
\end{equation} 
The Hamiltonian of the system can be expressed now in terms of 
the momentum conjugate to $\beta$ and the Lagrangian.
\begin{equation}
\label{504}
H_{dyn}  =  p_\beta \dot\beta - L \nonumber = 
{2\pi \over 3V}\, N\, {\rm e}^{3\Omega} p_\beta^2 + {3V\over 8\pi}\, 
{1\over N}\, \dot\Omega^2 {\rm e}^{-3\Omega}. 
\end{equation} 
The expression for the action integral 
\begin{equation} 
\label{505} 
I = \int p_\beta d\beta - H_{dyn} dt = \int p_\beta d\beta - 
\left[ {2\pi \over 3V}\, N\, {\rm e}^{3\Omega} p_\beta^2 + {3V\over 8\pi}\, 
{1\over N}\, \dot\Omega^2 {\rm e}^{-3\Omega}\right] dt 
\end{equation} 
differs in its appearance from the standard one. In particular, lapse $N$ 
is not a mere factor in front of some expression anymore. However, it is 
easy to see that variation of the action with respect to $N$ produces 
the constraint 
\begin{equation}
\label{506} 
p_\beta^2 = \left(  {3V \over 4\pi}\right)^2\, {1\over N^2}\, 
{\rm e}^{-6\Omega} \dot\Omega^2.
\end{equation}
that coincides with the Hamiltonian constraint of the standard approach 
when the variables of two different approaches are properly identified.  

Just as in previous section, we use the Hamilton--Jacobi equation (\ref{406}) 
together with the expression (\ref{504}) for the Hamiltonian $H_{dyn}$, 
to obtain the Schr\"odinger equation for the axisymmetric Kasner model.
\begin{equation}
\label{507}
i\hbar {\partial\Psi\over\partial t} = -{2\pi\hbar^2 \over 3V}\, N\,   
{\rm e}^{3\Omega}\, {\partial^2 \Psi \over \partial\beta^2} + 
{3V\over 8\pi}\, \dot\Omega^2\, {1\over N}\, {\rm e}^{-3\Omega}\, \Psi .
\end{equation}  
To complete the dynamic picture, this equation should be amended  by the 
equation 
\begin{equation}
\label{508}
<p_\beta>_s^2 = \left( {4\pi \over 3V}\right)^2\, {1\over N^2}\, 
{\rm e}^{-6\Omega}\,\dot\Omega^2.
\end{equation}
obtained from the constraint equation (\ref{506}) via the same procedure 
as equation (\ref{408}) of the previous section. 

The solution of these equations describing ``propagation'' of a wave packet 
of Kasner universes (cf. the previous section) can be written down right away. 
A simple inspection of equations (\ref{507}), (\ref{508}) reveals that they 
are obtained from the similar equations (\ref{407}), (\ref{408}) by replacing 
everywhere the factor ${\rm e}^{3\Omega}$ for the factor 
$N\, {\rm e}^{3\Omega}$. The solution of the Schr\"odinger equation 
(\ref{506}) is expressed by (\ref{419a}) with $f(t)$ given by 
\begin{equation}
\label{509}
f(t) = {2\pi\over 3V}\int\limits_{t_0}^t N\, {\rm e}^{3\Omega} dt, 
\end{equation} 
instead of that given by (\ref{423}). Correction of the equation (\ref{429}) 
is just as trivial. 
   
Up until now the lapse function $N(t)$ and, together with it, the meaning of 
the time parameter has not been specified. A variety of approaches can 
be used to fix both of them. For a case when shift is fixed (as in our 
example), one more condition including $N(t)$ should be imposed. One extreme 
case has been considered in the 
previous section. The lapse has been fixed and, after this, the meaning of 
the time parameter has been interpreted. Another possibility is to fix the 
interpretation of the time parameter, in which case the lapse function will 
be fixed implicitly and will be determined by an additional condition 
considered together with the dynamic evolution equations. 

As an example we consider the choice of $K = {\rm Tr}K$ as the time 
parameter. This choice might provide advantages in particular problems 
(such as avoidance of singularities, etc.) but we are not concerned with 
this now. In this case equations (\ref{507}) -- (\ref{509}) take the form   
\begin{equation}
\label{510}
i\hbar {\partial\Psi\over\partial K} = -{2\pi\hbar^2 \over 3V}\, N\,   
{\rm e}^{3\Omega}\, {\partial^2 \Psi \over \partial\beta^2} + 
{3V\over 8\pi}\, \left({d\Omega\over dK}\right)^2\, {1\over N}\, 
{\rm e}^{-3\Omega}\, \Psi 
\end{equation}  
\begin{equation}
\label{511}
<p_\beta>_s^2 = \left( {4\pi \over 3V}\right)^2\, {1\over N^2}\, 
{\rm e}^{-6\Omega}\,\left({d\Omega\over dK}\right)^2.
\end{equation} 
and 
\begin{equation}
\label{512}
f(K) = {2\pi\over 3V}\int\limits_{K_0}^K N(K)\, {\rm e}^{3\Omega (K)} dK, 
\end{equation} 
After solving the Schr\"odinger equation and computing the expectation 
value $<p_\beta>_s$, (\ref{511}) yields 
\begin{equation}
\label{513}
k_0^2 = \left( {3 V \over 4 \pi}\right)^2\, {1\over N^2}\, 
{\rm e}^{-6\Omega}\,\left({d\Omega\over dK}\right)^2.
\end{equation} 
which, together with the standard expression (this expression holds for any 
slicing parametrization, including parametrization by $K$)  
\begin{equation} 
\label{514} 
K = -\frac{3}{N}\, \frac{d\Omega}{dK} 
\end{equation} 
provides the basis for computing both $\Omega (K)$ and $N(K)$. 
Indeed, (\ref{514}) implies 
\begin{equation} 
\label{515} 
\frac{1}{N}\, \frac{d\Omega}{dK} = -\frac{K}{3} 
\end{equation} 
substitution of which in (\ref{513}) yields the equation for $\Omega (K)$ 
\begin{equation}
\label{516}
k_0^2 = {V^2 \over 16\pi^2}\, K^2\, {\rm e}^{-6\Omega}
\end{equation} 
It is remarkable that for this particular time parametrization the equation 
(\ref{516}) for $\Omega (K)$ is an algebraic equation (generically it should 
be differential). This equation implies 
\begin{equation} 
\label{517} 
\Omega (K) = \frac{1}{6}\, \ln\left(\frac{V^2}{16\pi^2}\, 
\frac{K^2}{k_0^2}\right) 
\end{equation} 
which, together with (\ref{514}), yields the expression for $N(K)$ 
\begin{equation} 
\label{518} 
N(K) = -\frac{3}{K}\, \frac{d\Omega}{dK} = -\frac{1}{K^2} 
\end{equation}

\section{Discussion.}
\label{V}

The issue of time in the standard approach to canonical quantum
gravity does not seem to have a satisfactory resolution no matter
which of the two quantization procedures is used. Both Dirac and ADM
quantization do not seem to be able to handle the emerging
difficulties of conceptual and technical nature. The difficulties can
be traced to identification of the entire 3--geometry of a spacelike
slicing as the principal dynamic object.  The resulting constrained
dynamics consists of proper dynamics that follows from variations of
dynamic variables, and of the constraints that enforce the fundamental
symmetries of gravitational systems (general covariance) and are
obtained by varying lapse and shift. The constraints essentially
restrict the evolution of gravitational systems to a ``shell''. The
peculiarity of gravitation is that on this shell all of the dynamics
is essentially determined by the constraints, thus rendering the
proper dynamic relations obsolete and unnecessary. This does not
present any problem in the classical theory. However, quantization of
such a classical theory is reduced to quantization of the constraint
equations. This essentially amounts to restricting the system states
to the shell and  neglecting ``off shell'' contributions, which leads to
seemingly intractable problems in describing the time evolution of
gravitational systems.

Our suggestion is to separate the true dynamics of gravity fields from
enforcing the symmetries. Practically, it is done via putting the
dynamic object in the geometrodynamic superspace of true gravitational
degrees of freedom and formulating of variational principles on this
superspace rather than on the superspace of total 3--geometries, while
leaving the embedding variables as free parameters reserved for
enforcing of the symmetries on the solutions of proper dynamic
equations at a suitable moment. The equations enforcing the symmetries
following from the general covariance are not obtained by variations
on the geometrodynamic superspace. They come from variations of the
action with respect to lapse and shift, as in standard
approach. Within the realm of the classical theory, the final outcome
of the theory predictions does not change (compared to the standard
approach).  Only the order of operations is changed. First, we
consider all the solutions of the proper dynamic equations (on and off
shell), and then force them on the shell via adjusting the embedding
parameters. Since the latter enter the solutions for true dynamic
variables, the final expressions for these variables, too, depend on
the outcome of forcing the system on the shell.  In the classical
domain, the resulting theory is indistinguishable from the standard
one.

The quantum theory, however, becomes quite different. We consider the
true geometrodynamics (on the superspace of true dynamic variables or
its phase space) as the object of quantization. The resulting equation
of the theory is a Schr\"odinger equation with the Hamiltonian that is
not a square root Hamiltonian and thus avoids most of the conceptual as
well as technical problems of time. The initial values as well as the
solutions of these equations contain embedding variables as functional
parameters. They do not know anything about the shell (one can say that
they include both on and off shell contributions). By themselves, they
cannot either violate or enforce constraints. This should be achieved
via adjusting embedding parameters.  In addition, the solutions contain lapse
and shift functions. These are responsible for the interpretation of
time. When imposing constraints, it is important to realize that they
might not be enforceable exactly (by simply treating them as
additional operator equations obtained by the same procedure as the
Schr\"odinger equation and applied to the state functional) if their
operator versions do not commute with the Hamiltonian of the
Schr\"odinger equation and cannot be made to commute via adjusting
embedding parameters.  It only means that, in this particular problem,
off shell contributions cannot be neglected. A weaker version of the
constraints should be introduced. The version chosen by us in examples
was to impose the constraints on expectations. In this way the
covariance requirement is satisfied as much as it can be satisfied. To
a reasonable extent, it is exactly satisfied in the sense that
predictions on the final hypersurface remain the same if the surface
is not changing. The question of the change of the prediction when the
surface is changing can be also answered positively by simple count of
equations and adjustable parameters.  The same is true, under reasonable 
conditions, about the multiple choice problem of time. 

As about interpretation of time (metric interpretation of the slicing 
parameter) our consideration indicates that, within our approach, it is 
equivalent to imposing one (or more, if shift is involved) additional 
relation that involves lapse (or lapse and shift), and will not create 
any problems if these additional relations do not involve true dynamic 
variables. We do not believe that even relations involving true dynamic 
variables can cause considerable difficulties if they are introduced 
in the way similar to the one described for constraints. In our examples 
the relations included only embedding variables, thus commuting with the 
Hamiltonian and not causing any troubles. It should be stressed, however, 
that all three components of the evolution description for quantum 
geometrodynamic system --- quantum dynamics itself, constraints enforcing 
the symmetries, and the interpretation of time --- emerge together as 
the solution of the total problem of geometrodynamic evolution.           
   
\section*{Acknowledgements}

For discussion, advice, or judgment on one or another issue taken up 
in this manuscript, we are indebted to S.~Chakrabarti, R.~Fulp, D.~Holz, 
L.~Norris, M.~Maclean, and J.~A.~Wheeler. One of us (AK) gratefully 
acknowledges the support of this research by the Theoretical Division 
of the LANL during the Summer, 1999.

\end{document}